\documentclass[12pt]{article}
\usepackage{epsfig}
\usepackage{amssymb}
\usepackage{amsmath}
\usepackage{amsfonts}
\usepackage{graphicx}
\usepackage{mathrsfs}
\DeclareMathAlphabet{\mathscrbf}{OMS}{mdugm}{b}{n}
\usepackage{mathabx}
\usepackage[dvips]{color}
\usepackage{multirow}
\usepackage{calc}
\usepackage{accents}

\makeatletter
\newcommand{\shorteq}{\mathrel{\mkern0.2mu\mathpalette\shorteq@\relax\mkern0.2mu}}
\newcommand{\shorteq@}[2]{\scalebox{0.5}[1]{$\m@th#1=$}}

\newcommand{\longeq}[1]{\mathrel{\mathpalette\longeq@{#1}}}
\newcommand{\longeq@}[2]{%
  \begingroup
  \sbox\z@{$\m@th#1=$}%
  \ifdim#2<\wd\z@
    \resizebox{#2}{\height}{\box\z@}%
  \else
    \ifdim#2<3\wd\z@
      \hbox to #2{$\m@th#1=\hss=\hss=\hss=$}%
    \else
      \hbox to #2{$\m@th#1=\cleaders\hbox to 0.2\wd\z@{\hss$#1=$\hss}\hfil=$}%
    \fi
  \fi
  \endgroup
}
\makeatother

\newcommand{\bsigma}{\boldsymbol{\sigma}}

\newcommand{\R}{\mathbb{R}}
\newcommand{\C}{\mathbb{C}}
\newcommand{\Z}{\mathbb{Z}}

\newcommand{\fu}{\mathfrak{u}}

\newcommand{\fn}{{\mathfrak{n}}}

\newcommand{\fL}{\mathfrak{L}}

\newcommand{\fT}{\mathfrak{T}}

\newcommand{\bk}{\mathbf{k}}

\newcommand{\bn}{{\mathbf{n}}}
\newcommand{\bfr}{\mathbf{r}}

\newcommand{\bx}{\mathbf{x}}
\newcommand{\by}{\mathbf{y}}

\newcommand{\bH}{\mathbf{H}}
\newcommand{\bI}{\mathbf{I}}

\newcommand{\bL}{\mathbf{L}}

\newcommand{\bM}{\mathbf{M}}

\newcommand{\bP}{\mathbf{P}}

\newcommand{\cA}{{\mathcal{A}}}
\newcommand{\cB}{\mathcal{B}}

\newcommand{\cE}{\mathcal{E}}

\newcommand{\cK}{\mathcal{K}}

\newcommand{\cP}{\mathcal{P}}

\newcommand{\cT}{\mathcal{T}}

\newcommand{\cX}{\mathcal{X}}

\newcommand{\be}{\begin{equation}}
\newcommand{\ee}{\end{equation}}
\newcommand{\bea}{\begin{eqnarray}}
\newcommand{\eea}{\end{eqnarray}}
\newcommand{\nn}{\nonumber}
\newcommand{\kt}{\rangle}
\newcommand{\br}{\langle}

\newcommand{\ed}{\end{document}}

\newcommand{\bi}{\begin{itemize}}
\newcommand{\ei}{\end{itemize}}

\newcommand{\bce}{\begin{center}}
\newcommand{\ece}{\end{center}}

\newcommand{\sE}{\mathscr{E}}

\newcommand{\sH}{\mathscr{H}}

\newcommand{\IM}{{\rm Im}}

\newcommand{\bcK}{{\boldsymbol{\cK}}}

\newcommand{\bigtau}{\mbox{\large$\tau$}}

\newcommand{\for}{{\mbox{\rm for}}}

\newcommand{\Square}{\mbox{\Large$\square$}}

\begin{document}

\title{{Pseudo-Hermiticity, Anti-Pseudo-Hermiticity, and Generalized Parity-Time-Reversal Symmetry\\ at  Exceptional Points}}

\author{Nil~\.{I}nce\thanks{E-mail address: nince19@ku.edu.tr}~, 
Hasan~Mermer\thanks{E-mail address: hmermer20@ku.edu.tr}~, and Ali~Mostafazadeh\thanks{Corresponding author, E-mail address:
amostafazadeh@ku.edu.tr}\\[6pt]
Departments of Physics and Mathematics, Ko\c{c} University,\\  34450 Sar{\i}yer,
Istanbul, T\"urkiye}

\date{ }
\maketitle

\begin{abstract}
For a diagonalizable linear operator $H:\sH\to\sH$ acting in a separable Hilbert space $\sH$, i.e., an operator with a purely point spectrum, eigenvalues with finite algebraic multiplicities, and a set of eigenvectors that form a Reisz basis of $\sH$, the pseudo-Hermiticity of $H$ is equivalent to its generalized parity-time-reversal ($\cP\cT$) symmetry, where the latter means the existence of an antilinear operator $\cX:\sH\to\sH$ satisfying $[\cX,H]=0$ and $\cX^2=1$. {The original proof of this result makes use of the anti-pesudo-Hermiticity of every diagonalizable operator $L:\sH\to\sH$, which means the existence of an antilinear Hermitian bijection \linebreak $\bigtau:\sH\to\sH$ satisfying $L^\dagger=\bigtau L\,\bigtau^{-1}$. We establish the validity of this result for block-diagonalizable operators},  i.e., those which have a purely point spectrum, eigenvalues with finite algebraic multiplicities, and a set of generalized eigenvectors that form a Jordan Reisz basis of $\sH$. {This allows us to  generalize the original proof of the equivalence of pseudo-Hermiticity and generalized $\cP\cT$-symmetry for diagonalizable operators to block-diagonalizable operators. For a pair of pseudo-Hermitian operators acting respectively in  two-dimensional and  infinite-dimensional Hilbert spaces, we obtain explicit expressions for the antlinear operators $\bigtau$ and $\cX$ that realize their anti-pseudo-Hermiticity and generalized $\cP\cT$-symmetry at and away from the exceptional points.}
\end{abstract}

\section{Introduction}

The term ``pseudo-Hermitian operator'' was initially used in the context of indefinite-metric quantum theories \cite{sudarshan-1960} to mean ``symmetric'' with respect to the indefinite inner product \cite{bognar} on the space of state vectors of these theories. Therefore, to decide if an operator is in this sense pseudo-Hermitian, one needs the information about this indefinite inner product. If the indefinite-inner-product space in question is obtained by endowing a Hilbert space $\sH$ with an indefinite metric operator $\eta_-$, one needs to know $\eta_-$ to make this decision.\footnote{An operator $\eta:\sH\to\sH$ acting in a Hilbert space $\sH$ is called a metric operator, if it is a Hermitian automorphism. Here and throughout this article by ``Hermitian'' we mean self-adjoint, and by ``automorphism'' we mean a linear bijection mapping $\sH$ onto $\sH$. Note that every Hermitian automorphism is bounded and has a bounded inverse \cite{reed-simon}.}  The current use of the term ``pseudo-Hermitian operator'' is based on a definition given in Ref.~\cite{p1} which does not involve the use of a particular indefinite inner product or metric operator. It identifies the pseudo-Hermiticity of a linear operator $H:\sH\to\sH$ with the requirement of the existence of a definite or indefinite metric operator $\eta$ satisfying,
	\be
	H^\dagger=\eta\,H\eta^{-1},
	\label{PH}
	\ee 
where $H^\dagger$ denotes the adjoint of $H$, \cite{p1}. This definition, naturally, leads to the questions of the existence and characterization of all definite and indefinite metric operators $\eta$ fulfilling (\ref{PH}) for a given linear operator $H$.
	
This notion of pseudo-Hermiticity was initially introduced in an attempt to provide a consistent mathematical framework for the study of non-Hermitian parity-time-reversal ($\cP\cT$) symmetric Hamiltonians. This was necessary for a reliable assessment of the claims regarding the utility of $\cP\cT$-symmetry in devising a genuine extension of quantum mechanics \cite{Bender-PRL-1998,Bender-JMP-1999}. Here and in what follows the term ``$\cP\cT$-symmetry'' of a linear operator $H$ acting in a separable Hilbert space $\sH$ refers to the condition,
	\be
	[H,\cP\cT]=0,
	\label{PT-sym}
	\ee
where $\cP$ and $\cT$ are respectively linear and antilinear Hermitian operators that act in $\sH$ and fulfill
	\begin{align}
	&\cP^2=1,
	&&\cT^2=1, &&[\cP,\cT]=0,
	\label{involution}
	\end{align}
and $0$ and $1$ stand for the zero and identity operators. An immediate consequence of (\ref{involution}) is that $\cP\cT$ is an antilinear involution, i.e., it is an antilinear operator satisfying
	\be
	(\cP\cT)^2=1.
	\label{PT-involution}
	\ee

Ref.~\cite{p1} provides a spectral representation of a large class of pseudo-Hermitian operators which involves the use of biorthonormal systems. The latter proved to be a powerful tool for deriving basic properties of pseudo-Hermitian operators \cite{p2,p3}. These in turn helped clarify the physical aspects of unitary quantum systems defined by $\cP\cT$-symmetric and pseudo-Hermitian Hamiltonians \cite{jpa-2004,cjp-2004}. Another important outcome of the study of pseudo-Hermitian operators is a solution of the notorious Hilbert-space problem in relativistic quantum mechanics of spin-0 and spin-1 fields \cite{ap-2006,jmp-2009,jmp-2017,Hawton-2019} and quantum cosmology \cite{cqg-2003,ap-2004}. 

The initial results on pseudo-Hermitian operators were confined to the class of ``diagonalizable operators,'' \cite{p1,p2,p3,jmp-2003,jpa-2003}. These are linear operators $H$ with a purely point spectrum\footnote{This means that the spectrum of $H$ consists only of eigenvalues \cite{Reed-Simon}.}, eigenvalues with finite algebraic multiplicities, and a set of eigenvectors $\psi_n$ that form a Reisz basis\footnote{A Reisz basis is a (Schauder) basis $\{\psi_n\}$ that is the image of an orthonormal basis $\{\epsilon_n\}$ under a bounded operator $B$ with bounded inverse, i.e., $\psi_n=B\epsilon_n$.} of the Hilbert space. This condition is equivalent to the existence of a complete bounded biorthonormal system $\{(\psi_n,\phi_n)\}$ where $\psi_n$ and $\phi_n$ are respectively eigenvectors of $H$ and $H^\dagger$, \cite{young,review}. 

Refs.~\cite{p3,jmp-2003} prove the following characterization theorem that elucidates the spectral properties of diagonalizable pseudo-Hermitian operators and their relevance to $\cP\cT$-symmetry.
	\begin{itemize}
	\item[]{\bf Theorem~1:} Let $H:\sH\to\sH$ be a diagonalizable linear operator acting in a separable Hilbert space $\sH$. Then the following conditions are equivalent.
	\begin{enumerate}
	\item $H$ is pseudo-Hermitian.
	\item Eigenvalues of $H$ are either real or come in complex-conjugate pairs, and   members of each pair have the same geometric multiplicities.
	\item $H$ commutes with an antilinear involution, i.e., there is an antilinear operator $\cX:\sH\to\sH$ such that $\cX^2=1$ and $[H,\cX]=0$.
	\end{enumerate}
	\end{itemize}
In view of Eqs.~(\ref{PT-sym}) and (\ref{PT-involution}), this theorem shows that for diagonalizable operators, $\cP\cT$-symmetry is a special case of pseudo-Hermiticity. 

Next, consider a general linear operator $H$ that satisfies Condition~3 of Theorem~1. Then we can define, $\tilde\cP:=\cX\cT$, for some antilinear (anti-)Hermitian operator $\cT$ satisfying $\cT^2=1$, and identify this condition with  $\tilde\cP\cT$-symmetry of $H$. It is important to note that $\tilde\cP$ may not commute with $\cT$ or satisfy $\tilde\cP^2=1$, but it does satisfy $(\tilde\cP\cT)^2=1$. For this reason, $\tilde\cP\cT$-symmetry  is a generalization of $\cP\cT$-symmetry, and we refer to Condition~3 of Theorem~1 as the  ``generalized $\cP\cT$-symmetry'' of $H$, \cite{jmp-2003,ijmpa-2006}.

An interesting mathematical phenomenon occurs when a perturbation of a linear operator obstructs its diagonalizability. In this case the geometric multiplicity of some of the eigenvalues of the operator undergoes an abrupt change at a particular value of the perturbation parameter. This is a special case of what is known as an exceptional point \cite{kato}. When a perturbation of a generic pseudo-Hermitian operator causes one or more complex-conjugate pair(s) of its eigenvalues to merge into a single real eigenvalue, the sum of their geometric multiplicities drops and an exceptional point emerges.\footnote{In the literature on $\cP\cT$-symmetric Hamiltonians, this scenario is often called ``spontaneous $\cP\cT$-symmetry breaking''.}

Exceptional points arise in a variety of physical systems \cite{Heiss-1990,Heiss-1998,Heiss-2001,berry-2004,Muller-2008,Heiss-2012} and have remarkable physical applications 
\cite{Doppler,Chen,Miri,Li} particularly in the context of systems possessing $\cP\cT$-symmetry \cite{Hodaei,Miri,Li} and pseudo-Hermiticity \cite{Xia,Xiong-2021,scipost-2022,Yin-2023,Hao-2023,Mondal}. This provides ample motivation for the study of the structure of pseudo-Hermitian operators and their relation to $\cP\cT$-symmetry at an exceptional point where the operator becomes non-diagonalizable \cite{p4, Scolarici,Sayyad-2023,Starkov,Montag}. 

Ref.~\cite{p4} reports some early developments in this direction. Its main result is 
the following partial extension of Theorem~1 to the class of block-diagonalizable operators. These are linear operators with a purely point spectrum such that their eigenvalues have finite algebraic multiplicities and there is Reisz basis of the Hilbert space consisting of generalized eigenvectors of $H$ that is a Jordan basis for $H$, \cite{axler}. The latter condition means that the matrix representation of $H$ in this basis is block-diagonal and each of its diagonal blocks has a Jordan canonical form. \pagebreak
	\begin{itemize}
	\item[]{\bf Theorem~2:} Let $H:\sH\to\sH$ be a block-diagonalizable operator acting in a separable Hilbert space $\sH$. Then the following conditions are equivalent.
	\begin{enumerate}
	\item $H$ is pseudo-Hermitian.
	\item Eigenvalues of $H$ are either real or come in complex-conjugate pairs, and  the geometric multiplicities and Jordan dimensions\footnote{Jordan dimensions are the sizes of the Jordan blocks corresponding to an eigenvalue.} corresponding to complex-conjugate pairs of eigenvalues are equal.
	\end{enumerate}
	\end{itemize}

For future reference, we wish to mention that the argument used in Ref.~\cite{p4} to show that Condition~1 of this theorem implies its Condition~2 applies also when $H$ satisfies 
	\be
	H^\dagger=\gamma\,H\,\gamma^{-1},
	\label{ph-gamma}
	\ee
for a general, possibly non-Hermitian, automorphism $\gamma:\sH\to\sH$. This observation and Theorem~2 prove the following theorem.
	\begin{itemize}
	\item[]{\bf Theorem~3:} A block-diagonalizable operator $H:\sH\to\sH$ acting in a separable Hilbert space $\sH$ is pseudo-Hermitian if and only of there is an automorphism $\gamma:\sH\to\sH$ such that $H^\dagger=\gamma\,H\,\gamma^{-1}$.
	\end{itemize}
	
The hypothesis of Theorem~2 holds trivially when $\sH$ is finite-dimensional. Because every finite-dimensional inner-product space is isomorphic to the complex Euclidean space $\C^N$, with $N$ being the dimension of $\sH$, we can reduce the discussion of the linear operators acting in $\sH$ to that of square matrices. Ref.~\cite{Zhang} shows that ``$\cP\cT$-symmetric'' matrices are pseudo-Hermitian. We can state this result in terms of a general finite-dimensional Hilbert space as follows.
	\begin{itemize}
	\item[]{\bf Theorem~4:} Let $H:\sH\to\sH$ be a block-diagonalizable linear operator acting in a finite-dimensional Hilbert space $\sH$, and $\cX$ be an antilinear bijection satisfying $[H,\cX]=0$. Then $H$ is pseudo-Hermitian.
	\end{itemize}
To clarify the connection between this theorem and the results of Ref.~\cite{Zhang}, let $\sH$ be $\C^{N\times 1}$ endowed with the Euclidean inner product, $\br \bx,\by\kt:=\bx^\dagger\by$, and let $H$ and $\cX$ be given by
	\begin{align}
	&H(\bx):=\bH\,\bx,
	&&\cX(\bx):=\bP\bx^*,
	\label{HX=}
	\end{align}
where $\C^{m\times n}$ stands for the vector space of $m\times n$ complex matrices, $\bx,\by\in\C^{N\times 1}$ are arbitrray, $\bx^\dagger$ marks the conjugate-transpose (Hermitian conjugate) of $\bx$, $\bH,\bP\in\C^{N\times N}$, and an asterisk denotes complex-conjugation. 

If we define, $\cP,\cT:\C^{N\times 1}\to\C^{N\times 1}$ by $\cP(\bx):=\bP\bx$ and $\cT(\bx):=\bx^*$, we have
	\begin{align}
	&\cX=\cP\cT, &&\cT^2=1,
	\label{XP2T2=}
	\end{align}
and $\cP\cT$-symmetry of the matrix $\bH$ corresponds to $[H,\cP\cT]=0$. Similarly pseudo-Hermiticity of $\bH$ means pseudo-Hermiticity of $H$. The authors of Ref.~\cite{Zhang} prove that $[H,\cP\cT]=0$ implies the pseudo-Hermiticity of $\bH$ (Theorem~3 of \cite{Zhang}). In their treatment they require $\bP$ to satisfy 
	\be
	\bP^2=\bI,
	\label{P2=}
	\ee 
where $\bI$ is the $m\times m$ identity matrix. Their proof of pseudo-Hermiticity of $\bH$, however, uses the weaker condition of the invertibility of $\bP$. To see this, suppose that $\bH$ and consequently $H$ are $\cP\cT$-symmetric, so that $[H,\cP\cT]=0$. Assuming that $\bP$ and consequently $\cP$ are invertible, and noting that $\cT$ is also invertible, we can write $[H,\cP\cT]=0$ in the form\footnote{By invertibility of $\cP$ and $\cT$, we mean that they are bijections.}
	\be
	\cT\,H\,\cT^{-1}=\cP^{-1}H\,\cP.
	\nn
	\ee
This equation together with the identity, $\cT^{-1}=\cT$, imply 
	\[\bH^*\bx=(\cT\,H\,\cT)\bx=(\cT\,H\,\cT^{-1})\bx=(\cP^{-1}H\,\cP)\bx=
	\bP^{-1}\bH\,\bP\bx.\]
Therefore $\bH^*=\bP^{-1}\bH\,\bP$. Combining this equation with the fact that a square matrix is pseudo-Hermitian if and only if it is similar to its complex-conjugate (Theorem~2 of Ref.~\cite{Zhang}) we establish the pseudo-Hermiticity of $\bH$. Nowhere in this argument do we use the condition, $\bP^2=\bI$. We only use the invertibility of $\bP$ which is equivalent to that of $\cP$ and $\cX$. This is also true for the proof of Theorem~3 of Ref.~\cite{Zhang}.

{Theorem~3 actually follows as a simple corollary of the following generalization of Theorems~1 and 2 which was proven in Ref.~\cite{SS-2003} several years before the publication of Ref.~\cite{Zhang}.} 
	\begin{itemize}
	\item[]{\bf Theorem~5:} Let $H:\sH\to\sH$ be a block-diagonal linear operator acting in a separable Hilbert space $\sH$. Then the following conditions are equivalent.
	\begin{enumerate}
	\item $H$ is pseudo-Hermitian.
	\item Eigenvalues of $H$ are either real or come in complex-conjugate pairs with members of each pair having the same geometric  multiplicities and Jordan dimensions.
	\item  $H$ commutes with an antilinear involution, i.e., there is an antilinear operator $\cX:\sH\to\sH$ such that $\cX^2=1$ and $[H,\cX]=0$.
	\end{enumerate}
	\end{itemize}
The following are useful observations:
	\begin{itemize}
	\item[-] The hypothesis of Theorem~5 holds if $\sH$ is finite-dimensional. In particular, it provides a characterization of pseudo-Hermiticity for matrices in terms of their generalized $\cP\cT$-symmetry.
	\item[-] Theorem~4 states that if $\sH$ is finite-dimensional, Condition~3 of Theorem~5 with the requirement ``$\cX^2=1$'' relaxed to ``$\cX$ is a bijection.'' implies Condition~1.
	\end{itemize}	

{The original proof of Theorem~1 which is given in \cite{p3} makes use of the curious fact that every block-diagonalizable operator $L:\sH\to\sH$ is anti-pseudo-Hermitian. This means that it satisfies $L^\dagger=\bigtau\,L\,\bigtau^{-1}$ for some antilinear Hermitian bijection $\bigtau:\sH\to\sH$. The purpose of the present article is to extend this result to the class of block-diagonalizable operators and use it to present a proof of Theorem~5 that is a direct extension of the original proof of Theorem~1.}

{The above notion of anti-pseudo-Hermiticity turns out to play an important role in understanding the origins of reciprocity in scattering theory \cite{Deak-Fulop,Sigwarth}. Consider the scattering problem defined by the stationary Schr\"odinger equation, $H\psi=k^2\psi$, where $H$ is a Hamiltonian operator of the form $H=H_0+V$ acting in a separable Hilbert space $\sH$, $H_0$ and $V$ are respectively the free Hamiltonian operator and the interaction potential, and $k$ is a wavenumber. Let $\bk$ be the wave vector for the incident wave, $\bfr$ be the position of a generic detector that placed at spatial infinity, $\fn:=k^{-1}\bk$, $\fn':=r^{-1}\bfr$, $r:=|\bfr|$, and $\psi_{\rm s}(\bn,\bn',r)$ be the asymptotic expression for scattered wave reaching this detector. Then reciprocity theorem states that $\psi_{\rm s}(\fn,\fn',r)=\psi_{\rm s}(\fn',\fn,r)$, \cite{Landau,Twersky1954}. Ref.~\cite{Deak-Fulop} gives a straightforward proof of this theorem for situations where $H_0$ commutes with an antilinear unitary (antiunitary) operator $\fu:\sH\to\sH$ satisfying
	\be
	V^\dagger=\fu\,V\,\fu^{-1}.
	\label{weak}
	\ee
It is important to note that the hypothesis of this theorem does not require $H$ to be Hermitian; reciprocity does not follow from unitarity.}

{A direct consequence of $[H_0,\fu]=0$ and Eq.~\eqref{weak} is $[H,\fu^2]=[V,\fu^2]=0$. This means that either $\fu^2$ represents a symmetry transformation of both $H_0$ and $H$ or $\fu^2$ is a constant multiple of the identity operator. In the latter (generic) case, antiunitarity of $\fu$ implies $\fu^2=\pm 1$. If $\fu^2=1$, $\fu^\dagger=\fu^{-1}=\fu$. Therefore, $\fu$ is an antilinear Hermitian bijection, and both $V$ and $H$ are anti-pseudo-Hermitian.}  

{In standard potential scattering in $d$ dimensions, where $\sH:=L^2(\R^d)$,  and $H_0$ and $V$ are respectively functions of the momentum and position operators, and $H_0$ is Hermitian, the above reciprocity theorem applies for both real as well as complex potentials. This is simply because $[H_0,\cT]=0$ and $V^\dagger=\cT\,V\,\cT^{-1}$, where $\cT:L^2(\R^d)\to L^2(\R^d)$ is the time-reversal operator given by $(\cT\psi)(\bx):=\psi(\bx)^*$, which is antiunitary. The fact that $\cT^2=1$ implies that $\cT$ is an anti-linear Hermitian bijection. Therefore $V$ and consequently $H$ are anti-pseudo-Hermitian operators. This shows that reciprocity in potential scattering follows from the time-reversal symmetry of the free Hamiltonian $H_0$ and the anti-pseudo-Hermiticity of the total Hamiltonian $H$ realized by $\cT$.}

\section{Spectral resolution of block-diagonalizable operators}

By definition, a linear operator $H:\sH\to\sH$ acting in a separable Hilbert space $\sH$ is called block-diagonalizable if it has a purely point spectrum, its eigenvalues have finite algebraic multiplicities, and there is a Reisz basis of $\sH$ consisting of generalized eigenvectors of $H$ that form a Jordan basis for $H$. 

Consider such a block-diagonalizable operator $H:\sH\to\sH$ and let $\cB$ denote a corresponding Jordan basis which is a Reisz basis of $\sH$. Let $n$ be the spectral label that marks distinct eigenvalues $E_n$ of $H$, and for each $l\in\Z^+$, $d_{n,l}:=\dim[{\rm ker}(H-E_n 1)^l]$, where ``ker'' abbreviates ``kernel'' or ``null-space'' of its argument \cite{axler}. Then the finiteness of the algebraic multiplicity of $E_n$ implies the existence of positive integers $q_n$ such that $d_{n,l}=d_{n,q_n}$ if an only if $l\geq q_n$. The numbers $d_{n,1}$ and $d_{n,q_n}$ are respectively called the geometric and algebraic multiplicities of $E_n$. Following Ref.~\cite{p4} we denote the former by $d_n$, i.e., $d_n:=d_{n,1}$. This is also called the ``degree of degeneracy'' of $E_n$. 
 
The span of generalized eigenvectors of $H$ associated with its eigenvalues $E_n$ form finite-dimensional invariant subspaces $\sH_n$ for $H$. These furnish a direct-sum decomposition of $\sH$; $\sH=\displaystyle\bigoplus_{n}\sH_n$. Let $H_n:\sH_n\to\sH_n$ be the restriction of $H$ to $\sH_n$ and $\cB_n:=\cB\cap\sH_n$. Then $\cB_n$ is a Jordan basis for $H_n$, and the matrix representation of $H_n$ in this basis is a block-diagonal matrix consisting of $d_n$ Jordan blocks. The latter have the form,
	\[\left[\begin{array}{cccccc}
	E_n&1&0&0&\cdots&0\\
	0& E_n&1&0&\cdots&0\\
	0& 0&E_n&1&\cdots&0\\
	\vdots&\ddots&\ddots&\ddots&\ddots&\vdots\\
	0&\cdots&0&0&E_n&1\\
	0&\cdots&0&0&0&E_n
	\end{array}\right].\]
Let $p_{n,a}$ denote the Jordan dimension associated with the spectral label $n$ and degeneracy label $a$, so that the corresponding Jordan block belongs to $\C^{p_{n,a}\times p_{n,a}}$. We can then denote the elements of $\cB_n$ by $|\psi_n,a,i\kt$, where $a\in\{1,2,\cdots,d_n\}$ and $i\in\{1,2,\cdots,p_{n,a}\}$. 

Next, we recall that $\cB$ is a Reisz basis of $\sH$. This means that there a bounded automorphism $A:\sH\to\sH$ with bounded inverse and an orthonormal basis $\cE$ of $\sH$ with elements $|\epsilon_n,a,i\kt$, such that
	\be
	|\psi_n,a,i\kt=A|\epsilon_n,a,i\kt.
	\label{psi=}
	\ee
Because $\cE$ is an orthonormal basis of $\sH$, 
	\begin{align}
	&\br\epsilon_m,a,i|\epsilon_n,b,j\kt=\delta_{mn}\delta_{ab}\delta_{ij},
	&&\sum_n\sum_{a=1}^{d_n}\sum_{i=1}^{p_{n,a}}
	| \epsilon_n,a,i\kt\br\epsilon_n,a,i|=1.
	\label{ortho}
	\end{align}
	
Let us define 
	\be
	|\phi_n,a,i\kt:=A^{-1\dagger}|\epsilon_n,a,i\kt.
	\label{phi=}
	\ee		
Then, we can use \eqref{psi=} -- \eqref{phi=} to infer that  $\{(|\psi_n,a,i\kt,|\phi_n,a,i\kt)\}$ is a bounded complete biorthonormal system of $\sH$, i.e., 
	\begin{align}
	&\br\phi_m,a,i|\psi_n,b,j\kt=\delta_{mn}\delta_{ab}\delta_{ij},
	&&\sum_n\sum_{a=1}^{d_n}\sum_{i=1}^{p_{n,a}}
	|\psi_n,a,i\kt\br\phi_n,a,i|=1.
	\label{biortho}
	\end{align}
Furthermore, because $\cB$ is a Jordan basis for $H$, we can express $H$ as \cite{p4},
	\begin{align}
	H&=A\,H_0A^{-1}=\sum_n\sum_{a=1}^{d_n}\Big(E_n\sum_{i=1}^{p_{n,a}}
	|\psi_n,a,i\kt\br\phi_n,a,i|+
	\sum_{i=1}^{p_{n,a}-1}|\psi_n,a,i\kt\br\phi_n,a,i+1|\Big),
	\label{spectral-rep-H}
	\end{align}
where 
	\begin{align}
	&H_0:=\sum_n\sum_{a=1}^{d_n}\Big(E_n\sum_{i=1}^{p_{n,a}}
	|\epsilon_n,a,i\kt\br\epsilon_n,a,i|+
	\sum_{i=1}^{p_{n,a}-1}|\epsilon_n,a,i\kt\br\epsilon_n,a,i+1|\Big).
	\label{spectral-rep-Hb}
	\end{align}
The latter is a block-diagonalizable operator with an orthonormal Jordan basis, namely $\cE$, \cite{Nagy-2015}.

A straightforward consequence of \eqref{biortho} and \eqref{spectral-rep-H} is that $|\phi_n,a,i\kt$ are generalized eigenvectors of $H^\dagger$ and that they form a Jordan basis for $H^\dagger$. It is also easy to see that $|\psi_n,a,1\kt$ and $|\phi_n,a,p_{n,a}\kt$  are respectively eigenvectors of $H$ and $H^\dagger$ with eigenvalues $E_n$ and $E_n^*$.

\section{Anti-pseudo-Hermiticity}

The proof of Theorem~1 that is given in Ref.~\cite{p3} rests on the fact that every diagonalizable operator is anti-pseudo-Hermitian. In this section we extend this result to the class of block-diagonalizable operators. First, we recall the definition of an anti-pseudo-Hermitian operator.
	\begin{itemize}
	\item[]{\bf Definition:} A densely-defined linear operator $L:\sH\to\sH$ acting in a separable Hilbert space $\sH$ is said to be anti-pseudo-Hermitian if there is a Hermitian antilinear bijection $\bigtau:\sH\to\sH$ satisfying $L^\dagger=\bigtau\,L\,\bigtau^{-1}$.
	\end{itemize}
Notice that the adjoint of an antilinear bijection $\fL:\sH\to\sH$ is the antilinear operator $\fL^\dagger$ that satisfies $\br\xi|\fL^\dagger\,\zeta\kt=\br\zeta|\fL\,\xi\kt$. In particular, $\fL$ is said to be Hermitian (respectively unitary or antiunitary) provided that $\br\xi|\fL\,\zeta\kt=\br\zeta|\fL\,\xi\kt$ (respectively $\br\fL\,\xi|\fL\,\zeta\kt=\br\zeta|\xi\kt$) for all $\xi,\zeta\in\sH$, \cite{wigner}.
 		
In the following, $H$ is a general block-diagonalizable operator acting in a separable Hilbert space $\sH$, $|\psi_n,a,i\kt$ and $|\phi_n,a,i\kt$ are respectively generalized eigenvectors of $H$ and $H^\dagger$ that form Jordan bases for $H$ and $H^\dagger$. They also form Reisz bases of $\sH$. $\cE$ is the orthonormal basis whose elements $|\epsilon_n,a,j\kt$ satisfy \eqref{psi=}, and $A:\sH\to\sH$ is a bounded automorphism with a bounded inverse which together with $|\psi_n,a,i\kt$ and $|\phi_n,a,i\kt$ satisfy \eqref{spectral-rep-H}.

	\begin{itemize}
	\item[]{\bf Lemma~1:} Let $S,\fT:\sH\to\sH$ be functions defined on $\sH$ by 
		\begin{align}
		S|\xi\kt&:=\sum_n\sum_{a=1}^{d_n}\sum_{i=1}^{p_{n,a}}
		\br\epsilon_n,a,i|\xi\kt~|\epsilon_n,a,p_{n,a}-i+1\kt,
		\label{S=}\\
		\fT |\xi\kt&:=\sum_n\sum_{a=1}^{d_n}\sum_{i=1}^{p_{n,a}}
		\br\xi|\epsilon_n,a,i\kt~|\epsilon_n,a,i\kt, 
		\label{Theta=}		
		\end{align}
	where $|\xi\kt\in\sH$ is arbitrary, and $\bigtau_0:=S\,\fT$. Then the following assertions hold.
		\begin{enumerate}
		\item $S$ is a Hermitian and unitary linear involution.
		\item $\fT$ is a Hermitian and unitary antilinear involution.
		\item $[S,\fT]=0$. 
		\item $\bigtau_0$ is a Hermitian and unitary antilinear involution.
		\end{enumerate}
	\item[]{\bf Proof:} $S$ is a linear operator, because the inner product $\br\,\cdot\,|\,\cdot\,\kt$ is linear in its second slot. $S$ is a unitary operator, because according to (\ref{S=}) it is a permutation of the basis vectors $|\epsilon_n,a,1\kt,|\epsilon_n,a,2\kt,\cdots,|\epsilon_n,a,p_{n,a}\kt$. It is an involution, because it swaps elements of disjoint pairs of these vectors. This shows that $S^2|\epsilon_n,a,i\kt=|\epsilon_n,a,i\kt$. Because $S$ is a linear operator, this implies $S^2=1$, i.e., $S$ is an involution. Combining this with the fact that $S$ is unitary, we have $S^\dagger=S^{-1}=S^{-1}S^2=S$. Therefore $S$ is Hermitian. This proves Assertion 1. $\fT$ is an antilinear operator, because  the inner product $\br\,\cdot\,|\,\cdot\,\kt$ is antilinear in its first slot. According to (\ref{Theta=}), 
		\be
		\fT |\epsilon_n,a,i\kt=|\epsilon_n,a,i\kt.
		\label{Theta-act}
		\ee
This equation together with (\ref{Theta=}) imply	
		\begin{align}
		\fT^2|\xi\kt&=\fT\Big(\sum_n\sum_{a=1}^{d_n}\sum_{i=1}^{p_{n,a}}
		\br\xi|\epsilon_n,a,i\kt~|\epsilon_n,a,i\kt\Big)\nn\\
		&=\sum_n\sum_{a=1}^{d_n}\sum_{i=1}^{p_{n,a}}
		\br\xi|\epsilon_n,a,i\kt^*|\epsilon_n,a,i\kt\nn\\
		&=\sum_n\sum_{a=1}^{d_n}\sum_{i=1}^{p_{n,a}}
		|\epsilon_n,a,i\kt\br\epsilon_n,a,i|\xi\kt\nn\\
		&=|\xi\kt.\nn
		\end{align}
Because this holds for all $|\xi\kt\in\sH$,  $\fT^2=1$. Another consequence of (\ref{Theta=}) is $\br\zeta|\fT|\xi\kt=\br\xi|\fT|\zeta\kt$ for all $|\xi\kt,|\zeta\kt\in\sH$. Because $\fT$ is anti-linear, this equation shows that it is a Hermitian, i.e., $\fT^\dagger=\fT$. Combining this equation and the fact that $\fT^2=1$, we find $\fT^\dagger=\fT=\fT^{-1}$. Therefore $\fT$ is unitary. This proves Assertion 2. In view of \eqref{S=}, \eqref{Theta=}, and (\ref{Theta-act}), for all $|\xi\kt\in\sH$,
	\begin{align}
	S\,\fT|\xi\kt&=\sum_n\sum_{a=1}^{d_n}\sum_{i=1}^{p_{n,a}}
		\br \xi|\epsilon_n,a,i\kt~S|\epsilon_n,a,i\kt,\nn\\
		&=\sum_n\sum_{a=1}^{d_n}\sum_{i=1}^{p_{n,a}}
		\br\epsilon_n,a,i|\xi\kt^*~|\epsilon_n,a,p_{n,a}-i+1\kt\nn\\
		&=\sum_n\sum_{a=1}^{d_n}\sum_{i=1}^{p_{n,a}}
		\br\epsilon_n,a,i|\xi\kt^*~\fT|\epsilon_n,a,p_{n,a}-i+1\kt\nn\\
		&=\fT\sum_n\sum_{a=1}^{d_n}\sum_{i=1}^{p_{n,a}}
		\br\epsilon_n,a,i|\xi\kt~|\epsilon_n,a,p_{n,a}-i+1\kt\nn\\
		&=\fT S|\xi\kt.\nn
	\end{align}
This proves Assertion 3. Because $\bigtau_0=S\,\fT$, the fact that $S$ and $\fT$ are respectively linear and antilinear unitary operators implies that $\bigtau_0$ is an antilinear unitary operator. Because $S^2=\fT^2=1$ and $[S,\fT]=0$, $\bigtau_0^2=(S\,\fT)^2=
S\,\fT S\,\fT=S^2\fT^2=1$. Therefore $\bigtau_0$ is an involution. Because $\bigtau_0^\dagger=\bigtau_0^{-1}=\bigtau_0$, $\bigtau_0$ is Hermitian. This proves Assertion 4.~$\Square$ 
 	\end{itemize}
It is sometimes useful to express the antilinear involution $\fT$ in the form \cite{jpa-2003},
	\be
	\fT=\sum_n\sum_{a=1}^{d_n}\sum_{i=1}^{p_{n,a}}|\epsilon_n,a,i\kt\star
		\br\epsilon_n,a,i|, 
		\label{Theta-star}	
	\ee
where $\star\br\epsilon_n,a,i|:\sH\to\C$ stands for the antilinear operator defined by,
	\[\star\br\epsilon_n,a,i|\xi\kt:=\br\epsilon_n,a,i|\xi\kt^*=\br\xi|\epsilon_n,a,i\kt.\]
	
\begin{itemize}
	\item[]{\bf Lemma~2:} Let $\bigtau_0$ be as in Lemma~1, and $H_0$ be the linear operator given by (\ref{spectral-rep-Hb}). Then 
	\be
	H_0^\dagger=\bigtau_0\,H_0\,\bigtau_0^{-1}.
	\label{tau-ph}
	\ee
In particular, $H_0$ is anti-pseudo-Hermitian.
	\item[]{\bf Proof:} Statement 2 of Lemma~1, Eq.~(\ref{Theta-act}), and the fact $\fT$ is antilinear imply
		\begin{align}
	\big(\fT|\epsilon_n,a,i\kt\br\epsilon_m,b,j|\fT\big)|\xi\kt&=
	\fT|\epsilon_n,a,i\kt\br\epsilon_m,b,j|\fT\,\xi\kt\nn\\
	&=\br\epsilon_m,b,j|\fT\xi\kt^*\fT|\epsilon_n,a,i\kt\nn\\
	&=\br\epsilon_m,b,j|\fT^\dagger\xi\kt^*|\epsilon_n,a,i\kt\nn\\
	&=\br\xi|\fT|\epsilon_m,b,j\kt^*|\epsilon_n,a,i\kt\nn\\
	&=\br\xi|\epsilon_m,b,j\kt^*|\epsilon_n,a,i\kt\nn\\
	&=|\epsilon_n,a,i\kt\br\epsilon_m,b,j|\xi\kt.\nn
		\end{align}
This proves
	\[\fT|\epsilon_n,a,i\kt\br\epsilon_m,b,j|\fT=
	|\epsilon_n,a,i\kt\br\epsilon_m,b,j|.\]
Making use of this identity, Eq.~(\ref{S=}), and Statement 1 of Lemma~1, we have
	\begin{align}
	\bigtau_0|\epsilon_n,a,i\kt\br\epsilon_m,b,j|\bigtau_0^{-1}&=
	S\,\fT|\epsilon_n,a,i\kt\br\epsilon_m,b,j|\fT^{-1}S^{-1}\nn\\
	&=S\,\fT|\epsilon_n,a,i\kt\br\epsilon_m,b,j|\fT S^\dagger\nn\\
	&=S|\epsilon_n,a,i\kt\br\epsilon_m,b,j|S^\dagger\nn\\
	&=|\epsilon_n,a,p_{n,a}-i+1\kt\br\epsilon_m,b,p_{m,b}-i+1|.
	\label{id1}
	\end{align}
With the help of this relation and \eqref{spectral-rep-Hb}, we can show that
	\begin{align}
	\bigtau_0\,H_0\,\bigtau_0^{-1}&=	
	\sum_n\sum_{a=1}^{d_n}\Big(E_n^*\sum_{i=1}^{p_{n,a}}
	\bigtau_0|\epsilon_n,a,i\kt\br\epsilon_n,a,i|\bigtau_0+
	\sum_{i=1}^{p_{n,a}-1}
	\bigtau_0|\epsilon_n,a,i\kt\br\epsilon_n,a,i+1|\bigtau_0 \Big),\nn\\
	&=\sum_n\sum_{a=1}^{d_n}\Big(E_n^*\sum_{i=1}^{p_{n,a}}
	|\epsilon_n,a,p_{n,a}-i+1\kt
	\br\epsilon_n,a,p_{n,a}-i+1|+\nn\\
	&\hspace{1.8cm}
	\sum_{i=1}^{p_{n,a}-1}
	|\epsilon_n,a,p_{n,a}-i+1\kt
	\br\epsilon_n,a,p_{n,a}-i|\Big)\nn\\
	&=\sum_n\sum_{a=1}^{d_n}\Big(E_n^*\sum_{j=1}^{p_{n,a}}
	|\epsilon_n,a,j\kt\br\epsilon_n,a,j|+\sum_{j=1}^{p_{n,a}-1}
	|\epsilon_n,a,j+1\kt\br\epsilon_n,a,j|\Big)\nn\\
	&=H_0^\dagger.~~\Square\nn
	\end{align}
\end{itemize}
	\begin{itemize}
	\item[]{\bf Theorem~6:} Every block-diagonalizable operator acting in a separable Hilbert space is anti-pseudo-Hermitian. 
	\item[]{\bf Proof:} Let $H$ be a block-diagonalizable operator acting in a separable Hilbert space $\sH$. Then as we show in Sec.~2, we can express $H$ in the form $H=A\,H_0A^{-1}$ where $H_0$ is given by (\ref{spectral-rep-Hb}). Let $\bigtau_0$ be the antilinear involution of Lemma~1, and 
		\be
		\bigtau:=(A\,\bigtau_0\,A^\dagger)^{-1}.
		\label{tau-def}
		\ee 
Then, $\bigtau$ is an antilinear Hermitian bijection, because $A$ and $\bigtau_0$ are respectively linear and antilinear bijections, and 
		\[\bigtau^\dagger=(A\,\bigtau_0\,A^\dagger)^{-1\dagger}=
		(A\,\bigtau_0^\dagger A^\dagger)^{-1}=
		(A\,\bigtau_0 A^\dagger)^{-1}=\bigtau.\]
Furthermore, in view of Lemma~2, we have
		\begin{align}
		H^\dagger&=A^{-1\dagger}H_0^\dagger\,A^\dagger=
		A^{-1\dagger}\bigtau_0\, H_0\,\bigtau_0^{-1}A^\dagger\nn\\
		&=A^{-1\dagger}\bigtau_0\, A^{-1}H\,A\,\bigtau_0^{-1}A^\dagger=
		A^{-1\dagger}\bigtau_0^{-1}\, A^{-1}H\,A\,\bigtau_0\,A^\dagger\nn\\
		&=\bigtau\,H\,\bigtau^{-1}.~~\Square
		\label{anti-ph=1}
		\end{align}
\end{itemize}

\section{Block-diagonalizable pseudo-Hermitian operators}

An immediate consequence of Theorem~6 is that every pseudo-Hermitian block-diagonalizable operator commutes with some antilinear bijection. This is because given a metric operator satisfying $H^\dagger=\eta\,H\,\eta^{-1}$, we can use (\ref{anti-ph=1}) to show that $H$ commutes with $\eta^{-1}\bigtau$ which is an antilinear bijection. For cases where $\sH$ is finite-dimensional, this provides a constructive proof of the converse of the statement of Theorem~4. In the following, we present a slightly stronger result (Theorem~7). First, we introduce some convenient notation and prove a useful lemma.

Theorem~2 states that if a block-diagonalizable operator $H$ is pseudo-Hermitian, its eigenvalues are either real or come in complex-conjugate pairs with identical geometric multiplicities and Jordan dimensions. To provide a quantitative demonstration of this property, we use $\nu_0$, $\nu$, and $-\nu$ instead of the spectral label $n$ whenever $E_n$ is real, has a positive imaginary part, and has a negative imaginary part, respectively, i.e.,
	\be
	n\longrightarrow\left\{
	\begin{array}{ccc}
	\nu_0 &\for&  \IM(E_n)=0,\\
	\nu &\for& \IM(E_n)>0,\\
	-\nu &\for&\IM(E_n)<0,\end{array}\right.
	\label{relabel}
	\ee
where ``$\IM(\cdot)$'' stand for the imaginary part of its argument. We can assume, without loss of generality, that $\nu_0$ and $\nu$ take positive integer values, because this is equivalent to changing $n$ to $-n$ when $\IM(E_n)<0$. 

In view of Theorem~2 and (\ref{relabel}), the pseudo-Hermiticity of $H$ implies,
	\begin{align} 	
	&E_{-\nu}=E_{\nu}^*,  
	&&d_{-\nu}=d_\nu, 
	&&p_{-\nu,a}=p_{\nu,a}.
	\nn
	\end{align}
This allows us to express Eq.~\eqref{spectral-rep-Hb} in the form
	\begin{align}
	H_0=&\sum_{\nu_0}\sum_{a=1}^{d_{\nu_0}}
	\Big(E_{\nu_0}\sum_{i=1}^{p_{{\nu_0},a}}
	|\epsilon_{\nu_0},a,i\kt\br\epsilon_{\nu_0},a,i|+
	\sum_{i=1}^{p_{{\nu_0},a}-1}
	|\epsilon_{\nu_0},a,i\kt\br\epsilon_{\nu_0},a,i+1|\Big)+\nn\\
	&\sum_{\nu}\sum_{a=1}^{d_{\nu}}\Big[\sum_{i=1}^{p_{{\nu},a}}
	\Big(E_{\nu}|\epsilon_{\nu},a,i\kt\br\epsilon_{\nu},a,i|+
	E_{\nu}^*|\epsilon_{-\nu},a,i\kt\br\epsilon_{-\nu},a,i|\Big)+\nn\\
	&\hspace{1.3cm}\sum_{i=1}^{p_{{\nu},a}-1}
	\big(|\epsilon_{\nu},a,i\kt\br\epsilon_{\nu},a,i+1|+
	|\epsilon_{-\nu},a,i\kt\br\epsilon_{-\nu},a,i+1|\big)\Big].
	\label{spectral-rep-Hb-ph}
	\end{align} 

	\begin{itemize}
	\item[]{\bf Lemma~3:}  Let $S$ and $H_0$ be respectively given by \eqref{S=} and (\ref{spectral-rep-Hb-ph}), $C_0,\eta_0:\sH\to\sH$ be the linear operators defined on $\sH$ by
	\be
	C_0:=\sum_{\nu_0}\sum_{a=1}^{d_{\nu_0}}\sum_{i=1}^{p_{\nu_0,a}}
	|\epsilon_{\nu_0},a,i\kt\br\epsilon_{\nu_0},a,i|+
	\sum_{\nu}\sum_{a=1}^{d_{\nu}}\sum_{i=1}^{p_{\nu,a}}\Big(
	|\epsilon_{-\nu},a,i\kt\br\epsilon_{\nu},a,i|+
	|\epsilon_{\nu},a,i\kt\br\epsilon_{-\nu},a,i|\Big),
	\label{Cb=}
	\ee
and $\eta_0:=SC_0$. Then $C_0$ and $\eta_0$ are linear Hermitian and unitary involutions.
	\item[]{\bf Proof:} The fact that $C_0^\dagger=C_0$ follows from \eqref{Cb=}. Because according to this equation $C_0$ does not change $|\epsilon_{\nu_0},a,i\kt$ but swaps $|\epsilon_{\nu},a,i\kt$ and $|\epsilon_{-\nu},a,i\kt$, $C_0^2=1$. This together with $C_0^\dagger=C_0$ imply $C_0^{-1}=C_0^\dagger$. Therefore, $C_0$ is a linear Hermitian and unitary involution. Because $S$ is unitary,
	\begin{align}
	SC_0S^{-1}=SC_0S^\dagger&=\sum_{\nu_0}\sum_{a=1}^{d_{\nu_0}}
	\sum_{i=1}^{p_{\nu_0,a}}
	|\epsilon_{\nu_0},a,p_{\nu_0,a}-i+1\kt\br\epsilon_{\nu_0},a,p_{\nu_0,a}-i+1|+\nn\\
	&\hspace{.5cm}\sum_{\nu}\sum_{a=1}^{d_{\nu}}\sum_{i=1}^{p_{\nu,a}}\Big(
	|\epsilon_{-\nu},a,p_{\nu,a}-i+1\kt\br\epsilon_{\nu},a,p_{\nu,a}-i+1|+\nn\\
	&\hspace{2.5cm}
	|\epsilon_{\nu},a,p_{\nu,a}-i+1\kt\br\epsilon_{-\nu},a,p_{\nu,a}-i+1|\Big)\nn\\
	&=C_0.\nn
	\end{align}
Therefore,
	\be
	[C_0,S]=0,
	\label{Cb-S}
	\ee 
and consequently 
	\[\eta_0^2=(SC_0)^2=SC_0SC_0=S^2C_0^2=1,\]
i.e., $\eta_0$ is an involution. Next, we note that because both $S$ and $C_0$ are unitary operators, so is $\eta_0$. In light of this observation and Eq.~\eqref{Cb-S}, we have 
	\[\eta_0^\dagger=\eta_0^{-1}=(SC_0)^{-1}=C_0^{-1}S^{-1}=C_0S=SC_0=\eta_0.\] 
Therefore, $\eta_0$ is Hermitian.~$\Square$
	\end{itemize}			
	\begin{itemize}
	\item[]{\bf Theorem~7:} Let $H$ be a block-diagonalizable pseudo-Hermitian operator acting in a separable Hilbert space $\sH$. Then there is an antilinear involution $\cX:\sH\to\sH$ that commutes with $H$,  i.e., $H$ possesses generalized $\cP\cT$-symmetry.
	\item[]{\bf Proof:} Let $C_0$ and $\eta_0$ be as in Lemma~3. To simplify some of the calculations, we switch back to using the original spectral label $n$ and introduce
	\[|\tilde\epsilon_n,a,i\kt:=C_0|\epsilon_n,a,i\kt.\] 
This implies
	\begin{align}
	&H_0|\tilde\epsilon_n,a,i\kt=E_n^*|\tilde\epsilon_n,a,i\kt,
	\label{H-Cb}\\
	&\eta_0|\epsilon_n,a,i\kt=SC_0|\epsilon_n,a,i\kt=
	S|\tilde\epsilon_n,a,i\kt=|\tilde\epsilon_n,a,p_{n,a}-i+1\kt,
	\label{eta-b}
	\end{align}
where we have employed (\ref{S=}). According to \eqref{spectral-rep-Hb} and \eqref{eta-b},
	\begin{align}
	\eta_0\,H_0\,\eta_0^{-1}&=
	\sum_n\sum_{a=1}^{d_n}\Big(E_n\sum_{i=1}^{p_{n,a}}
	\eta_0|\epsilon_n,a,i\kt\br\epsilon_n,a,i|\eta_0^\dagger+
	\sum_{i=1}^{p_{n,a}-1}\eta_0|\epsilon_n,a,i\kt\br\epsilon_n,a,i+1|\eta_0^\dagger\Big),\nn\\
	&=\sum_n\sum_{a=1}^{d_n}\Big(E_n\sum_{i=1}^{p_{n,a}}
	|\tilde\epsilon_n,a,p_{n,a}-i+1\kt\br\tilde\epsilon_n,a,p_{n,a}-i+1|+\nn\\
	&\hspace{2cm}
	\sum_{i=1}^{p_{n,a}-1}
	|\tilde\epsilon_n,a,p_{n,a}-i+1\kt\br\tilde\epsilon_n,a,p_{n,a}-i|\Big)\nn\\
	&=\sum_n\sum_{a=1}^{d_n}\Big(E_n\sum_{j=1}^{p_{n,a}}
	|\tilde\epsilon_n,a,j\kt\br\tilde\epsilon_n,a,j|+\sum_{j=1}^{p_{n,a}-1}
	|\tilde\epsilon_n,a,j+1\kt\br\tilde\epsilon_n,a,j|\Big)\nn\\
	&=\sum_n\sum_{a=1}^{d_n}\Big(E_n^*\sum_{j=1}^{p_{n,a}}
	|\epsilon_n,a,j\kt\br\epsilon_n,a,j|+\sum_{j=1}^{p_{n,a}-1}
	|\epsilon_n,a,j+1\kt\br\epsilon_n,a,j|\Big)\nn\\
	&=H_0^\dagger.\nn
	\end{align}
	Combining the latter equation with (\ref{tau-ph}) and setting $\cX_0:=\eta_0^{-1}\bigtau_0=\eta_0\bigtau_0$, where $\bigtau_0$ is the antilinear involution of Lemma~1, we see that $\cX_0$ is an antilinear operator satisfying
	\be
	[H_0,\cX_0]=0.
	\label{X-b}
	\ee
Next, we note that  by virtue of \eqref{eta-b},
	\be
	\eta_0=\sum_n\sum_{a=1}^{d_n}\sum_{i=1}^{p_{n,a}}
	|\tilde\epsilon_n,a,p_{n,a}-i+1\kt\br\epsilon_n,a,i|.
	\label{eta-b=}
	\ee
This equation together with (\ref{id1}), \eqref{Cb-S}, and the fact that $S$ is a Hermitian involution imply
	\[\bigtau_0\,\eta_0\,\bigtau_0^{-1}=S\,\eta_0\,S^\dagger=S^2C_0 S=SC_0=\eta_0.\]
Therefore $[\eta_0,\bigtau_0]=0$ and
$\cX_0^2=(\eta_0\bigtau_0)^2=\eta_0^2\bigtau_0^2=1$.
Finally, we let 
	\be
	\cX:=A\,\cX_0A^{-1}=A\,\eta_0\bigtau_0A^{-1}=A\,\eta_0\,S\,\fT\, A^{-1}.
	\label{cX-def=}
	\ee 
Because $A$ and $\cX_0$ are respectively linear and antilinear operators, $\cX$ is an antilinear operator. Furthermore, we have $\cX^2=A\cX_0^2A^{-1}=1$ and 
	\begin{align} 
	&[H,\cX]=[AH_0A^{-1},A\cX_0A^{-1}]=A[H_0,\cX_0]A^{-1}=0.~~\Square
	\end{align}
	\end{itemize}
	
Next, we give an infinite-dimensional extension of Theorem~4. 
\begin{itemize}
	\item[]{\bf Theorem~8:} Let $H:\sH\to\sH$ be a block-diagonalizable linear operator acting in a (possibly infinite-dimensional) separable Hilbert space $\sH$, and $\cX$ be an antilinear bijection satisfying $[H,\cX]=0$. Then $H$ is pseudo-Hermitian.
	\item[]{\bf Proof:} Because $\cX$ is a bijection, we can express $[H,\cX]=0$ as 
	\be
	H=\cX\, H\,\cX^{-1}.
	\label{H=XHX}
	\ee
Next, we recall that according to Theorem~6, there is an antilinear bijection $\bigtau:\sH\to\sH$ satisfying $H^\dagger=\bigtau\,H\,\bigtau^{-1}$. Substituting (\ref{H=XHX}) in this equation, we find
	\be
	H^\dagger=\gamma\, H\,\gamma^{-1},
	\label{gamma}
	\ee
where $\gamma:=\bigtau\,\cX$. Because $\bigtau$ and $\cX$ are antilinear bijections, $\gamma$ is an automorphism. Therefore, according to Theorem~3, (\ref{gamma}) implies that $H$ is pseudo-Hermitian.~~$\Square$
\end{itemize}

\begin{itemize}
\item[]{\bf Proof of Theorem~5:} The equivalence of Conditions 1 and 2 of Theorem~5 is the subject of Theorem~2 which is proven in Ref.~\cite{p4}. Therefore, it suffices to prove the equivalence of Conditions 1 and 3. Theorem~7 shows that Condition~1 implies Condition~3. Because every antilinear involution is a bijection, Theorem~8 implies the converse.~~$\Square$
\end{itemize}

\section{Examples}

\subsection{Pseudo-Hermiticity in real potential scattering}

Stationary scattering admits a dynamical formulation in which the scattering data can be extracted from the evolution operator for an effective non-unitary quantum system \cite{ap-2014,pra-2021,pra-2023}. For real potential scattering in one dimension, the latter is a two-level system defined by a matrix Hamiltonian of the form,
	\be
	{H}(x)=\frac{v(x)}{2k}\left[\begin{array}{cc}
	1 & e^{-2ikx}\\
	-e^{2ikx} & -1\end{array}\right]=\frac{v(x)}{2k}\,
	e^{-ikx\bsigma_3}\bcK\, e^{ikx\bsigma_3},
	\label{bH=}
	\ee
where $v:\R\to\R$ is a short-range potential\footnote{This means that $\int_{-\infty}^\infty |v(x)|dx$ exists and $\lim_{x\to\pm\infty} x^{1+\alpha} v(x)=0$ for some positive real number $\alpha$.}, $k$ is a positive real parameter corresponding to the wavenumber of the incident plane wave, and
	\begin{align}
	&\bsigma_3:=\left[\begin{array}{cc}
	1 & 0\\
	0& -1\end{array}\right],
	&&\bcK:=\left[\begin{array}{cc}
	1 & 1\\
	-1& -1\end{array}\right].
	\label{bcK=}
	\end{align}
It turns out that the transfer matrix $\bM$ of the potential $v(x)$ which determines its reflection and transmission amplitudes is given in term of the time-evolution operator for ${H}(x)$ where $x$ plays the role of time. Specifically, 
	\[\bM=U(+\infty,-\infty):=\lim_{x_\pm\to\pm\infty} U(x_+,x_-),\]
where for all $x,x_0\in\R$, $U(x,x_0)$ is the solution of 
	\begin{align}
	&i\partial_x U(x,x_0)={H}(x) U(x,x_0),
	&& U(x_0,x_0)=\bI.\nn
	\end{align}

Because both the trace and determinant of $\bcK$ vanish, this matrix and consequently $H(x)$ fail to be diagonalizable. It is also easy to check that $\bcK^\dagger=\bsigma_3\,\bcK\,\bsigma_3$ which implies ${H}(x)^\dagger=\bsigma_3{H}(x)\,\bsigma_3^{-1}$. Therefore, ${H}(x)$ is a non-diagonalizable pseudo-Hermitian operator.\footnote{We arrive at the same conclusion, by noting that zero is the only eigenvalue of ${H}(x)$ and using Theorem~2.} In the following, we give the explicit form of the mathematical objects we have employed in the preceding sections for ${H}(x)$. In particular, we obtain the corresponding generalized $\cP\cT$-symmetry operator $\cX$.

We begin our analysis by recalling that for this example, $\sH$ is obtained by endowing $\C^{2\times 1}$ with the Euclidean inner product, $\br\bx|\by\kt:=\bx^\dagger\by$. This allows us to identity linear operators $L:\sH\to\sH$ with $2\times 2$ matrices $\bL$ satisfying $L\bx=\bL\bx$. 

Because zero is the only eigenvalue of ${H}(x)$, we have $n=\nu_0=1$, $d_1=1$, and $p_{1,1}=2$. Furthermore, we can identify the generalized eigenvectors of ${H}(x)$ that form a Jordan basis with 
	\begin{align}
	&|\psi_1,1,1\kt:=e^{-ikx\bsigma_3}\left[\begin{array}{c}
	1 \\ -1\end{array}\right]=A|\epsilon_1,1,1\kt,	
	&&|\psi_1,1,2\kt:=e^{-ikx\bsigma_3}\left[\begin{array}{c}
	1 \\ 0\end{array}\right]=A|\epsilon_1,1,2\kt,
	\label{psi=1}
	\end{align}
where
	\begin{align}
	&A:=e^{-ikx\bsigma_3}\left[\begin{array}{cc}
	1 & 1\\
	-1 & 0\end{array}\right],
	&&|\epsilon_1,1,1\kt:=\left[\begin{array}{c}
	1 \\ 0\end{array}\right],
	&&|\epsilon_1,1,2\kt:=\left[\begin{array}{c}
	0 \\ 1\end{array}\right],
	\label{A=}
	\end{align}
The biorthonormal complements of $|\psi_1,1,i\kt$ and the operator $H_0$ of Eq.~\eqref{spectral-rep-Hb} have the form,
	 \begin{align}
	&|\phi_1,1,1\kt:=e^{-ikx\bsigma_3}\left[\begin{array}{c}
	0 \\ 1\end{array}\right],	
	\quad\quad\quad|\phi_1,1,2\kt:=e^{-ikx\bsigma_3}\left[\begin{array}{c}
	1 \\ 1\end{array}\right],
	\label{phi=1}\\
	&H_0=|\epsilon_1,1,1\kt\br\epsilon_1,1,2|=\left[\begin{array}{cc}
	0 & 1 \\
	0 & 0\end{array}\right].
	\label{H0=1}
	\end{align}
	
Next, we compute the operators $S,\fT,\bigtau_0,\bigtau,C_0,\eta_0,\cX_0$, and $\cX$ of Lemmas~1 and 3, and Theorems~6 and 7. Using \eqref{S=}, \eqref{Theta=} and recalling \eqref{tau-def}, \eqref{Cb=}, \eqref{cX-def=}, $\bigtau_0:=S\,\fT$,  $\eta_0:=SC_0$, and $\cX_0:=\eta_0\bigtau_0$, we obtain
	\begin{align}
	&S=\eta_0=\bsigma_1:=\left[\begin{array}{cc}
	0 & 1\\
	1 & 0
	\end{array}\right],
	&& \fT=\cX_0=\cT,
	&& \bigtau_0=\bsigma_1\,\cT,\nn\\
	&\bigtau=
	\left[\begin{array}{cc}
	e^{-2ikx} & 0\\
	-1 & -e^{2ikx}\end{array}\right]\cT,
	&& C_0=\bI, 
	&&  \cX=e^{-2ikx\bsigma_3}\cT.\nn
	\end{align}
where $\cT$ stands for complex-conjugation, $\cT\bx:=\bx^*$. This calculation allows for an explicit demonstration of the anti-pseudo-Hermiticity of the Hamiltonian operator (\ref{bH=}) and the fact that it possess generalized $\cP\cT$-symmetry generated by an antilinear involution $\cX$ of the form $\tilde\cP\cT$ with $\tilde\cP:=e^{-2ikx\bsigma_3}$. Notice that $\tilde\cP$ is a linear unitary operator that fails to be an involution or commute with $\cT$ unless $k x$ is a half-integer multiple of $\pi$, in which case $\tilde\cP=\pm\bI$ and $\cX=\pm\cT$.

\subsection{A pseudo-Hermitian operator acting in an infinite-dimensional Hilbert space}

Let $\sH_0$ be a separable Hilbert space, $\Lambda:\sH_0\to\sH_0$ be a bounded Hermitian operator with a purely point spectrum and non-degenerate eigenvalues which are necessarily real. We label the latter by $\lambda_\ell$ with $\ell\in\Z^+$ such that $\lambda_{\ell+1}>\lambda_\ell$ for all $\ell\in\Z^+$. Suppose that $\lambda_1>0$. Then because $\Lambda$ is bounded, we have
	\[0<\lambda_1<\lambda_2<\lambda_3<\cdots<\parallel\Lambda\parallel,\]
where $\parallel\Lambda\parallel$ is the operator (Sup.) norm of $\Lambda$.

Let $\sH:=\C^{2\times 1}\otimes \sH_0$ and $H:\sH\to\sH$ be the linear operator defined on $\sH$ by
	\be
	H:=\frac{1}{2\varpi}\left(\Lambda\,\bcK-2\varpi^2\bsigma_3\right)=
	\frac{1}{2\varpi}\left[\begin{array}{cc}
	\Lambda-2\varpi^2 & \Lambda\\
	-\Lambda & -\Lambda+2\varpi^2\end{array}\right],
	\label{H=eq3}
	\ee
where $\varpi$ is a real and positive parameter, and $\bcK$ and $\bsigma_3$ are the matrices given by \eqref{bcK=}. Again, because  $\bcK^\dagger=\bsigma_3\,\bcK\,\bsigma_3$ and $\bsigma_3^2=\bI$, we have $H^\dagger= \bsigma_3^{-1}H\,\bsigma_3$. Therefore $H$ is pseudo-Hermitian. This operator is a simplified version of an effective Hamiltonian operator that is used in Ref.~\cite{scipost-2022} to study exceptional points associated with a class of real scattering potentials in two dimensions. In the following we apply the machinery developed in the preceding sections to determine the antilinear involution $\cX$, i.e., the generalized $\cP\cT$-symmetry operator for $H$. 

It is not difficult to show that $H$ is diagonalizable unless if $\varpi=\sqrt{\lambda_{\ell_\star}}$ for some $\ell_\star\in\Z^+$. These values of $\varpi$ mark exceptional points of order 2. In the following we use $\sE$ to denote the set of these exceptional points, i.e., $\sE:=\big\{\sqrt\lambda_\ell\:\big|\:\ell\in\Z^+\big\}$. In the following we consider the cases where $\varpi\notin \sE$ and $\varpi\in \sE$ separately.
	
If $\varpi\notin \sE$, eigenvalues of $H$ and a corresponding set of its eigenvectors are as follows.
	\begin{align}
	&E_{\ell\pm}:=\pm E_\ell,
	&&|\psi_{\ell\pm}\kt:=\frac{1}{2\varpi}\left[\begin{array}{c}
	\varpi\mp E_\ell\\
	\varpi\pm E_\ell\end{array}\right]|\lambda_\ell\kt,
	\end{align}
where $E_\ell:=\sqrt{\varpi^2-\lambda_\ell}$, and $|\lambda_\ell\kt$ are eigenvectors of $\Lambda$ with eigenvalue $\lambda_\ell$ which form an orthonormal basis of $\sH_0$, i.e.,  
	\begin{align}
	&\Lambda|\lambda_\ell\kt=\lambda_\ell|\lambda_\ell\kt,
	&&\br\lambda_k|\lambda_\ell\kt=\delta_{k\ell},
	\label{ortho}\\
	&\sum_{\ell=1}^\infty |\lambda_\ell\kt\br\lambda_\ell|=1,
	&&
	\sum_{\ell=1}^\infty \lambda_\ell\,|\lambda_\ell\kt\br\lambda_\ell|=\Lambda.
	\label{spec-res-Lambda}
	\end{align} 
Note that because $\varpi$ and $\lambda_\ell$ take real values, the eigenvalues of $H$ are either real or come in complex-conjugate (imaginary) pairs. If $\varpi<\sqrt{\lambda_1}$ (respectively $\varpi>\sqrt{\parallel\Lambda\parallel}$), none of them are real (respectively imaginary). If for some $m\in\Z^+$, $\sqrt{\lambda_m}<\varpi<\sqrt{\lambda_{m+1}}$, there are $2m$ real eigenvalues ($m$ pairs of real eigenvalues having opposite signs) and infinitely many complex-conjugate imaginary eigenvalues.
	
We can express $|\psi_{\ell\pm}\kt$ in terms of the orthonormal basis $\cE$ of $\sH$ consisting of the vectors,
	\begin{align}
	&|\epsilon_{\ell+}\kt:=\left[\begin{array}{c}
	1\\
	0\end{array}\right]|\lambda_\ell\kt,
	&&|\epsilon_{\ell-}\kt:=\left[\begin{array}{c}
	0\\
	1\end{array}\right]|\lambda_\ell\kt.
	\label{epsilons}
	\end{align}
Specifically, we have
	\begin{align}
	&|\psi_{\ell\pm}\kt=A|\epsilon_{\ell\pm}\kt,
	&&A:=\frac{1}{2}\left[\begin{array}{cc}
	1-\varpi^{-1}E&1+\varpi^{-1}E\\
	1+\varpi^{-1}E& 1-\varpi^{-1}E\end{array}\right],
	\label{A=eg2}
	\end{align}
where
	\[E:=\sum_{\ell=1}^\infty E_\ell\,|\lambda_\ell\kt\br\lambda_\ell|=
	\sum_{\ell=1}^\infty\sqrt{\varpi^2-\lambda_\ell}~|\lambda_\ell\kt\br\lambda_\ell|=\sqrt{\varpi^2-\Lambda}.\]
Notice that $E$ is a normal operator with real or complex-conjugate imaginary pairs of non-degenerate eigenvalues. Furthermore, it is bounded and has a bounded inverse. This in turn implies that the same holds for $A$. Therefore, $|\psi_{\ell\pm}\kt$ form a Reisz basis of $\sH$, and $H$ is a diagonalizable operator. 

Calculating the inverse of $A$, we find
	\be
	A^{-1}=\frac{1}{2}\left[\begin{array}{cc}
	1-\varpi\,E^{-1}&1+\varpi\,E^{-1}\\
	1+\varpi\,E^{-1}& 1-\varpi\,E^{-1}\end{array}\right].
	\label{inv-A=eg2}
	\ee
With the help of this equation, we derive the following expressions for the biorthonormal complement of $|\psi_{n\pm}\kt$ and the spectral resolution of $H$, respectively.
	\begin{align}
	&|\phi_{\ell\pm}\kt=A^{-1\dagger}|\epsilon_{\ell\pm}\kt=\frac{1}{2E_\ell^*}
	\left[\begin{array}{c}
	E_\ell^*\mp\varpi\\
	E_\ell^*\pm\varpi\end{array}\right]|\lambda_\ell\kt,\\
	&H=\sum_{\ell=1}^\infty E_\ell\left(|\psi_{\ell+}\kt\br\phi_{\ell+}|-
	|\psi_{\ell-}\kt\br\phi_{\ell-}|\right)=A\,H_0\,A^{-1},
	\end{align} 
where
	\be
	H_0:=\sum_{\ell=1}^\infty E_\ell\left(|\epsilon_{\ell+}\kt\br\epsilon_{\ell+}|-
	|\epsilon_{\ell-}\kt\br\epsilon_{\ell-}|\right)=E\,\bsigma_3.
	\label{H0-eg2}
	\ee

Next, suppose that $\varpi\in\sE$, i.e., $\varpi=\sqrt{\lambda_{\ell_\star}}$ for some $\ell_\star\in\Z^+$. In this case, $E_{\ell_\star\pm}=E_{\ell_\star}=0$, and
	\[|\psi_{\ell_\star-}\kt=|\psi_{\ell_\star+}\kt=\frac{1}{2}\left[\begin{array}{c}
	1\\
	1\end{array}\right]|\lambda_{\ell_\star}\kt=\frac{1}{2}\left(|\epsilon_{\ell_\star-}\kt+
	|\epsilon_{\ell_\star+}\kt\right).\]
Therefore, we need to supplement the set of the eigenvectors $|\psi_{\ell\pm}\kt$ with an addition vector to form a basis of $\sH$. It is easy to show that $\sH_{\ell_\star}$ is spanned by $|\psi_{\ell_\star+}\kt$ and the following generalized eigenvector of $H$ that is associated with its null eigenvalue. 
	\be
	|\psi_{\ell_\star+},1,2\kt=-\frac{1}{\sqrt{\lambda_{\ell_\star}}}\left[\begin{array}{c}
	1 \\ 0\end{array}\right]=-\frac{1}{\sqrt{\lambda_{\ell_\star}}}\:|\epsilon_{\ell_\star+}\kt.
	\ee
Here we note that $d_{\ell_\star}=1$, $p_{\ell_\star,1}=2$, and use the notation of the preceding sections. To conform with this notation, we denote $|\psi_{\ell_\star+}\kt$ by $|\psi_{\ell_\star+},1,1\kt$ and use $\cB$ to specify the basis consisting of $|\psi_{\ell_\star+},1,j\kt$ with $j\in\{1,2\}$ and $|\psi_{\ell\pm}\kt$ with $\ell\in\Z^+\setminus\{\ell_\star\}$.

To show that $\cB$ is a Reisz basis of $\sH$, we first construct a linear operator $A_\star$ that maps it onto $\cE$. This is given by
	\begin{align}
	A_\star&:=\cA_{\ell_\star}\Pi_{\ell_\star}+A\,(1-\Pi_{\ell_\star}),
	\label{A-star-def}
	\end{align}
where
	\begin{align}	
	&\cA_{\ell_\star}:=\frac{1}{2}\left[\begin{array}{cc}
	1 & -2/\sqrt{\lambda_{\ell_\star}}\\
	1 & 0\end{array}\right],
	&&\Pi_{\ell_\star}:=|\lambda_{\ell_\star}\kt\br\lambda_{\ell_\star}|,\nn
	\end{align}
and $A$ is the operator defined in (\ref{A=eg2}). It is not difficult to show that $A_\star$ is a bounded operator with a bounded inverse, namely
	\[A_\star^{-1}=\cA_{\ell_\star}^{-1}\Pi_{\ell_\star}+A^{-1}(1-\Pi_{\ell_\star}).\]
This shows that $\cB$ is a Reisz basis of $\sH$. 

Having calculated $A_\star^{-1}$, we can use it to construct the biorthonormal complements of $|\psi_{\ell_\star+},1,1\kt$ and $|\psi_{\ell_\star+},1,2\kt$. The result is 
	\begin{align}
	&|\phi_{\ell_\star+},1,1\kt:=A_\star^{-1\dagger}|\epsilon_{\ell_\star+}\kt
	=\cA_{\ell_\star}^{-1\dagger}|\epsilon_{\ell_\star+}\kt=
	\left[\begin{array}{c}
	0\\
	2\end{array}\right]|\lambda_{\ell_\star}\kt=2 |\epsilon_{\ell_\star-}\kt,\nn\\
	&|\phi_{\ell_\star+},1,2\kt:=A_\star^{-1\dagger}|\epsilon_{\ell_\star-}\kt
	=\cA_{\ell_\star}^{-1\dagger}|\epsilon_{\ell_\star-}\kt=
	\sqrt{\lambda_{\ell_\star}}\left[\begin{array}{c}
	-1\\
	1\end{array}\right]|\lambda_{\ell_\star}\kt=\sqrt{\lambda_{\ell_\star}}
	\left(|\epsilon_{\ell_\star-}\kt-|\epsilon_{\ell_\star+}\kt\right).\nn
	\end{align}
The spectral resolution of $H$ given in terms of the elements of $\cB$ and its biorthonormal complement has the form,
	\be
	H=|\psi_{\ell_\star+},1,1\kt\br\phi_{\ell_\star+},1,2|+ 
	\sum_{\ell=1}^\infty E_\ell\left(|\psi_{\ell+}\kt\br\phi_{\ell+}|-
	|\psi_{\ell-}\kt\br\phi_{\ell-}|\right)=A_\star\,H_0\,A_\star^{-1},
	\label{H-star=}
	\ee
where we have made use of the fact that $E_{\ell_\star}=0$, and introduced
	\begin{align}
	H_0&:=|\epsilon_{\ell_\star+}\kt\br\epsilon_{\ell_\star-}|+ 
	\sum_{\ell=1}^\infty E_\ell\left(|\epsilon_{\ell+}\kt\br\epsilon_{\ell+}|-|\epsilon_{\ell-}\kt\br\epsilon_{\ell-}|\right)\nn\\
	&=\left[\begin{array}{cc}
	0 & 1\\
	0 & 0\end{array}\right]\Pi_{\ell_\star}+\bsigma_3\,E.
	\label{H0-eg-2-star}
	\end{align}
In view of \eqref{H-star=} and (\ref{H0-eg-2-star}), $\cB$ is a Jordan basis for $H$. 

Next, we compute the antilinear involution $\fT$. It is easy to see that according to \eqref{Theta-star} and \eqref{epsilons},
	\be
	\fT=\sum_{\ell=1}^\infty\left(|\epsilon_{\ell+}\kt\star\br\epsilon_{\ell+}|+
	|\epsilon_{\ell-}\kt\star\br\epsilon_{\ell-}|\right)
	=\sum_{\ell=1}^\infty|\lambda_\ell\kt\star\br\lambda_\ell|.
	\label{fT=eg2}
	\ee
    The calculation of $S,\bigtau_0,\bigtau,C_0,\eta_0,\cX_0$, and $\cX$ requires separate analysis for the cases where $\varpi\in\sE$ and $\varpi\notin\sE$. We give the details of this calculation in the appendix. Here we give the final result of this calculation for {the antilinear Hermitian bijection $\bigtau$ which establishes the anti-pseudo-Hermiticity of $H$ and} the generalized $\cP\cT$-symmetry operator $\cX$:
	\begin{align}
	&{\bigtau=\left\{
	\begin{aligned} 
	&A^{-2}\sum_{\ell=1}^\infty|\lambda_\ell\kt\star\br\lambda_\ell|
	&& {\rm if}~ \varpi\in\sE,\\
	&A^{-2}\!\!\!\!\! \sum_{\ell=1,\, \ell \neq \ell_\star}^\infty|\lambda_\ell\kt\star\br\lambda_\ell|-2\sqrt{\lambda_{\ell_\star}}
	\left[\begin{array}{cc} 0 & 1\\
	1 & -2\end{array}\right]|\lambda_{\ell_\star}\kt\star\br\lambda_{\ell_\star}|
	&& {\rm if}~ \varpi=\sqrt{\lambda_{\ell_\star}}\notin\sE,
	\end{aligned}\right.}\nn\\[12pt]
	&\cX=\left\{
	\begin{aligned} 
	&\bsigma_1\fT&&{\rm if}~ 
	\varpi<\sqrt{\lambda_1},\\ 
	&\Big[(1-\bsigma_1)\sum_{\ell=1}^m|\lambda_\ell\kt\br\lambda_\ell|+\bsigma_1\Big]\fT
	&&{\rm if}~\sqrt{\lambda_m}\leq\varpi<\sqrt{\lambda_{m+1}}~\mbox{for some $m\in\Z^+$},\\ 
	&\fT&&{\rm if}~ 
	\varpi>\sqrt{\parallel\Lambda\parallel}.
	\end{aligned}\right.\nn
	\end{align}
A surprising outcome of this calculation is that $\cX_0$ coincides with $\cX$. Note also that
	\[\Big[(1-\bsigma_1)\sum_{\ell=1}^m|\lambda_\ell\kt\br\lambda_\ell|+\bsigma_1\Big]\fT=\sum_{\ell=1}^m|\lambda_\ell\kt\star\br\lambda_\ell|+\bsigma_1
	\sum_{\ell=m+1}^\infty|\lambda_\ell\kt\star\br\lambda_\ell|.\]

\section{Concluding remarks}

The discovery of Schr\"odinger operators involving complex $\cP\cT$-symmetric potentials with real spectra and the early claims regarding their possible role in generalizing quantum mechanics in the late 1990's \cite{Bender-PRL-1998,Bender-JMP-1999} led to an extensive study of the behavior of these operators. Among early developments in this direction is the introduction of the notion of a pseudo-Hermitician operator as defined in Ref.~\cite{p1}. This provided a suitable framework for examining the utility of $\cP\cT$-symmetry in generalizing quantum mechanics \cite{review}. It also found applications in dealing with the old problem of constructing Hilbert spaces in relativistic quantum mechanics and quantum cosmology \cite{ap-2006,jmp-2017,ap-2004}. 

An important outcome of the study of the relationship between pseudo-Hermiticity and $\cP\cT$-symmetry is the discovery of the equivalence of pseudo-Hermiticity of a diagonalizable operator and the existence of an antilinear involution $\cX$ that commutes with it (generalized $\cP\cT$-symmetry) \cite{p3}. This in particular identifies $\cP\cT$-symmetry as a special case of pseudo-Hermiticity for diagonalizable operators. {The original proof of the equivalence of pseudo-Hermiticity and generalized $\cP\cT$-symmetry for diagonalizable operator makes use of the curious fact that every diagonalizable operator is anti-pseudo-Hermitian. In the present article, we showed that this assertion holds for general block-diagonalizable operators and used it to obtain a direct extension of the original proof of the above-mentioned equivalence between pseudo-Hermiticity and generalized $\cP\cT$-symmetry to the class of block-diagonalizable operators.} 

An important aspect of the present investigation is that the proofs of its results are constructive. {This allows us to obtain various linear and antilinear operators appearing in our analysis. In particular, we can construct anitilinear Hermitian bijections $\bigtau$ that achieve the anti-pseudo-Hermiticity of a given block-diagonalizable operator and the antilinear involution that realizes the generalized $\cP\cT$-symmetry of a pseudo-Hermitian operator at its exceptional points or away from them.}\vspace{12pt}


\section*{Appendix: $S,\bigtau_0,\bigtau,C_0,\eta_0,\cX_0$, and $\cX$ for the operator \eqref{H=eq3}}

The calculation of $S,\bigtau_0,\bigtau,C_0,\eta_0,\cX_0$, and $\cX$ for the operator $H$ that is given by \eqref{H=eq3} requires separate analyses for the cases where $\varpi\in\sE$ and $\varpi\notin\sE$.

If $\varpi\notin\sE$, $H$ is diagonalizable, and \eqref{S=}, \eqref{Theta-star}, \eqref{tau-def}, \eqref{A=eg2}, \eqref{inv-A=eg2}, and $\bigtau_0=S\fT$ give
	\begin{align}
	&S=1,
	&&\bigtau_0=\fT=\sum_{\ell=1}^\infty
	|\lambda_\ell\kt\star\br\lambda_\ell|,
	&&\bigtau=A^{-1\dagger 2}\,\fT=A^{-2}\,\fT,
	\label{S-fT=eg2}
	\end{align}
and
	\[A^{-2}=\frac{1}{2}\left[\begin{array}{cc}
	1+\varpi^2 E^{-2} & 1-\varpi^2 E^{-2}\\
	1-\varpi^2 E^{-2}&1+\varpi^2 E^{-2}\end{array}\right].\]
To determine $C_0,\eta_0,\cX_0$, and $\cX$, we consider the following possibilities separately.
\begin{itemize}
\item[$\bullet$] $\varpi<\sqrt{\lambda_1}\,$: Then eigenvalues of $H$ are imaginary, and in view of \eqref{tau-def}, \eqref{Cb=}, \eqref{cX-def=}, \eqref{spec-res-Lambda}, \eqref{epsilons}, \eqref{A=eg2}, \eqref{inv-A=eg2}, \eqref{S-fT=eg2}, $\eta_0:=SC_0$, and $\cX_0:=\eta_0\bigtau_0$, we have
	\begin{align}
	&\eta_0=C_0=\sum_{l=1}^\infty \left(|\epsilon_{\ell+}\rangle \langle \epsilon_{\ell-}| + |\epsilon_{\ell-}\rangle \langle \epsilon_{\ell+}| \right)=\bsigma_1,\nn\\
	&\cX_0=\cX=\bsigma_1\fT=\bsigma_1\sum_{\ell=1}^\infty|\lambda_\ell\kt\star\br\lambda_\ell|,\nn
	\end{align}
where $\bsigma_1$ is the first Pauli matrix, i.e.,
	\[\bsigma_1:=\left[\begin{array}{cc}
	0 & 1\\
	1 & 0\end{array}\right],\]
and we have benefitted from the identity $A\,\bsigma_1A^{-1}=\bsigma_1$.
\item[$\bullet$] $\varpi>\sqrt{\parallel\Lambda\parallel}\,$: Then eigenvalues of $H$ are real, and similar calculations give
	\begin{align}
	&\eta_0=C_0=1,
	&&\cX_0=\cX=\fT=\sum_{\ell=1}^\infty|\lambda_\ell\kt\star\br\lambda_\ell|.\nn
	\end{align}	
\item[$\bullet$] $\sqrt{\lambda_m}<\varpi<\sqrt{\lambda_{m+1}}$ for some $m\in\Z^+$: Then $H$ has $2m$ real eigenvalues, and we have
	\begin{align}
	&\begin{aligned}
	\eta_0=C_0&=\sum_{\ell=1}^m \left( |\epsilon_{\ell+}\rangle \langle \epsilon_{\ell+}|
	+ |\epsilon_{\ell-}\rangle \langle \epsilon_{\ell-}| \right)
	+ \sum_{\ell=m+1}^\infty \left( |\epsilon_{\ell+}\rangle \langle \epsilon_{\ell-}| +
	|\epsilon_{\ell-}\rangle \langle \epsilon_{\ell+}| \right)\nn\\
	&=(1-\bsigma_1)\sum_{\ell=1}^m|\lambda_\ell\kt\br\lambda_\ell|+\bsigma_1,
	\end{aligned}\\
	&\cX_0=\cX=\Big[(1-\bsigma_1)\sum_{\ell=1}^m|\lambda_\ell\kt\br\lambda_\ell|+\bsigma_1\Big]\fT=\sum_{\ell=1}^m|\lambda_\ell\kt\star\br\lambda_\ell|+\bsigma_1
	\sum_{\ell=m+1}^\infty|\lambda_\ell\kt\star\br\lambda_\ell|.
	\nn
	\end{align}
\end{itemize}

Finally, we consider the cases where $\varpi\in\sE$, i.e., $\varpi=\sqrt{\lambda_{\ell_\star}}$ for some $\ell_\star\in\Z^+$, and $H$ is not diagonalizable but block-diagonalizable. Then, according to \eqref{S=}, \eqref{Theta-star},  \eqref{spec-res-Lambda}, \eqref{epsilons}, \eqref{A=eg2}, \eqref{inv-A=eg2},
\eqref{fT=eg2},  $\bigtau_0=S\fT$, $\eta_0:=SC_0$, $\cX_0:=\eta_0\bigtau_0$, and $\cX:=A_\star\cX_0 A_\star^{-1}$,
	\begin{align}
	S&=|\epsilon_{\ell_\star+}\rangle \langle \epsilon_{\ell_\star -}| + |\epsilon_{\ell_\star -}\rangle \langle \epsilon_{\ell_\star +}| + \sum_{\ell=1,\, l \neq \ell_\star}^\infty \left( |\epsilon_{\ell+}\rangle \langle \epsilon_{\ell+}| + |\epsilon_{\ell-}\rangle \langle \epsilon_{\ell-}| \right)=1+(\bsigma_1-\bI)\Pi_{\ell_\star},\nn\\
	\bigtau_0&=
	\fT+(\bsigma_1-\bI)|\lambda_{\ell_\star}\kt\star\br\lambda_{\ell_\star}|
	=\sum_{\ell=1,\, \ell \neq \ell_\star}^\infty|\lambda_\ell\kt\star\br\lambda_\ell|+ \bsigma_1|\lambda_{\ell_\star}\kt\star\br\lambda_{\ell_\star}|,\nn\\
	\bigtau&=A^{-2}(1-\Pi_{\ell_\star})\fT-2\sqrt{\lambda_{\ell_\star}}
	\left[\begin{array}{cc} 0 & 1\\
	1 & -2\end{array}\right]\Pi_{\ell_\star}\fT,\nn\\
	&=
	A^{-2}\sum_{\ell=1,\, \ell \neq \ell_\star}^\infty|\lambda_\ell\kt\star\br\lambda_\ell|-2\sqrt{\lambda_{\ell_\star}}
	\left[\begin{array}{cc} 0 & 1\\
	1 & -2\end{array}\right]|\lambda_{\ell_\star}\kt\star\br\lambda_{\ell_\star}|,\nn
	\end{align}
	\begin{align}
	C_0&=\sum_{\ell=1}^{\ell_\star} \left( |\epsilon_{\ell+}\rangle \langle \epsilon_{\ell+}| + |\epsilon_{\ell-}\rangle \langle \epsilon_{\ell-}| \right)
+ \sum_{\ell=\ell_\star+1}^\infty \left( |\epsilon_{\ell+}\rangle \langle \epsilon_{\ell-}| + |\epsilon_{\ell-}\rangle \langle \epsilon_{\ell+}| \right)\nn\\
	&=\sum_{\ell=1}^{\ell_\star}|\lambda_\ell\kt\br\lambda_\ell|+
	\bsigma_1\sum_{\ell=\ell_\star+1}^{\infty}|\lambda_\ell\kt\br\lambda_\ell|=
	(\bI-\bsigma_1)\sum_{\ell=1}^{\ell_\star}|\lambda_\ell\kt\br\lambda_\ell|+\bsigma_1,	\nn\\
	\eta_0&=(\bI-\bsigma_1)\sum_{\ell=1}^{\ell_\star-1}|\lambda_\ell\kt\br\lambda_\ell|+\bsigma_1,\nn\\
	\cX_0&=\cX=\Big[(\bI-\bsigma_1)\sum_{\ell=1}^{\ell_\star}|\lambda_\ell\kt
	\br\lambda_\ell|+\bsigma_1\Big]\fT=
	\sum_{\ell=1}^{\ell_\star}|\lambda_\ell\kt\star\br\lambda_\ell|+\bsigma_1
	\sum_{\ell=\ell_\star+1}^{\infty}|\lambda_\ell\kt\star\br\lambda_\ell|.
	\nn
	\end{align}
 \vspace{12pt}
 
 \noindent {\bf Note:} {The authors were unaware of Ref.~\cite{SS-2003} when they prepared the first draft of the present article. The approach pursued in Ref~\cite{SS-2003} towards proving their Theorem~3 (which is equivalent to our Theorems~5, 7, and 8) differs from ours in that it does not make use of the notion of anti-pseudo-Hermiticity. {It shares the constructive nature of our analysis, as it also makes of the framework provided in Ref.~\cite{p4}.}}
  \vspace{12pt}

\noindent {\bf Data Availability:} No datasets were generated or analyzed during the current study.
\vspace{12pt}

\noindent {\bf Acknowledgements:}
This work has been supported by T\"urkiye Bilimler Akademisi 
(Turkish Academy of Sciences).

\ed